\documentclass[aps,prb,twocolumn,showpacs,groupedaddress]{revtex4-1}  
\usepackage{graphicx}  
\usepackage{dcolumn}   
\usepackage{bm}        
\usepackage{amssymb}   
\usepackage{amsmath}
\usepackage{color}
\begin{document}



\title{Unveiling the scattering behavior of small spheres}
\date{\today}
\author{Dimitrios C. Tzarouchis}

\email{dimitrios.tzarouchis@aalto.fi}
\affiliation{%
 Department of Radio Science and Engineering, Aalto University,\\ 
 P.O. Box 13000, FI-00076 Aalto, Finland}
 
\author{Pasi Yl\"{a}-Oijala}

\affiliation{%
 Department of Radio Science and Engineering, Aalto University,\\ 
 P.O. Box 13000, FI-00076 Aalto, Finland}
\author{Ari Sihvola}%

\email{ari.sihvola@aalto.fi}
\affiliation{%
 Department of Radio Science and Engineering, Aalto University,\\ 
 P.O. Box 13000, FI-00076 Aalto, Finland}

\begin{abstract}

{A classical way for exploring the scattering behavior of a small sphere is to approximate Mie coefficients with a Taylor series expansion. This ansatz delivered a plethora of insightful results, mostly for small spheres supporting electric localized plasmonic resonances. However, many scattering aspects are still uncharted, especially with regards to magnetic resonances. Here, an alternative system ansatz is proposed based on the Pad\'{e} approximants for the Mie coefficients. The results reveal the existence of a self-regulating radiative damping mechanism for the first magnetic resonance and new general resonating aspects for the higher order multipoles. Hence, a systematic way of exploring the scattering response is introduced, sharpening our understanding about the sphere's scattering behavior and its emergent functionalities.}

%
\end{abstract}

\maketitle


Light scattering and absorption from a single homogeneous sphere is a widely studied canonical problem encountered in branches such as material physics, chemistry, nanotechnology, and engineering~\cite{Fan2014}. New fundamental phenomena about the scattering and absorptive behavior of a sphere were recently understood, such as the anomalous light scattering~\cite{Tribelsky2006} or the identification of Fano-like resonant line-shapes of the scattering spectrum~\cite{Lukyanchuk2010}. Additionally, many more novel functionalities emerged through the metamaterial paradigm~\cite{Jahani2016,Geffrin2012} reinforcing in this way its long standing significance. 

A first, mathematically rigorous attempt at finding a physically sound explanation for the triggered scattering mechanisms was derived by Lord Rayleigh for very small scatterers (electrostatic case)~\cite{rayleigh1871scattering}. Later developments attributed to Thomson, Love, Lorenz, Debye and Mie~\cite{kerker2013scattering} delivered a full electrodynamic perspective about this problem. Lorenz--Mie (simply Mie) coefficients rigorously quantified the material and size contributions of the overall scattering behavior as a set of fractional functions consisting of spherical Bessel, Hankel, and Riccati--Bessel functions, viz., 
\begin{equation}\label{an}
 a_n=\frac{m^2 j_n(mx)\left[xj_n(x)\right]'-\mu_c j_n(x)\left[mxj_n(mx)\right]'}{m^2 j_n(mx)\left[xh_n^{(1)}(x)\right]'-\mu_c h_n^{(1)}(x)\left[mxj_n(mx)\right]'}
\end{equation}
\begin{equation}\label{bn}
 b_n=\frac{\mu_c j_n(mx)\left[xj_n(x)\right]'-j_n(x)\left[mxj_n(mx)\right]'}{\mu_c j_n(mx)\left[xh_n^{(1)}(x)\right]'-h_n^{(1)}(x)\left[mxj_n(mx)\right]'}
\end{equation}
where $a_n$ and $b_n$ denotes the electric and magnetic coefficients, respectively~\cite{bohren2008absorption}. The size parameter, $x=ka$, is a function of the sphere's radius $a$ and host medium wavenumber $k=\omega\sqrt{\varepsilon\mu}$; $\varepsilon_1$ and $\mu_1$ are the sphere's material parameters with a wavenumber of $k_1=\omega\sqrt{\varepsilon_1\mu_1}$. Finally, $m=\frac{k_1}{k}$ is the contrast parameter.

This blend of straightforward and complicated expressions rarely offers any physical intuition on the studied problem. A widely used method in circumventing the aforementioned obstacle is to approximate Eq.~(\ref{an}) and~(\ref{bn}) with a Taylor series expansion for small $x$, i.e.,

\begin{equation}\label{a1Taylor2}
 a_1^T\approx-i\frac{2}{3}\frac{\varepsilon_c - 1}{\varepsilon_c +2 }x^3-i\frac{2}{5}\frac{(\varepsilon_c-2)(\varepsilon_c-1)}{(\varepsilon_c +2 )^2}x^5+O[x]^6
\end{equation}

\begin{equation}\label{b1Taylor2}
 b_1^T\approx-i\frac{1}{45}(\varepsilon_c-1)x^5+O[x]^7
\end{equation}

where $\varepsilon_c=\varepsilon_1/\varepsilon$ is the permittivity a magnetically inert ($\mu_c=1$) dielectric sphere~\cite{bohren2008absorption,kerker2013scattering,stratton2007electromagnetic}. 

The first term of expression~\ref{a1Taylor2} is often characterized as a \emph{static} term~\cite{sihvola2006character}, while higher order terms as \emph{dynamic depolarization} terms~\cite{Meier1983}. Indeed, the aforementioned system ansatz offers important physical insights and intuition about the sphere's scattering features~\cite{capolino2009theory} mostly due to its form (Eq.~(\ref{a1Taylor2})) (or its inverse form $(a_1^T)^{-1}$), which allows a clear perspective about the material induced resonances; useful results have been extracted regarding small spheres/scatterers dipole behavior~\cite{sihvola2006character,capolino2009theory}, mainly for the electric resonances triggered by the localized surface plasmon (plasmonic) oscillations~\cite{Tribelsky2006,bohren2008absorption}.

However, the aforementioned system ansatz cannot be easily applied for studying the magnetic Mie terms, since its Taylor series expansion converges slowly with respect to the size parameter $x$ for the first magnetic resonance (Eq.~(\ref{b1Taylor2})). This inherent characteristic can be somehow improved by including higher order terms, resulting in long and complicated expressions. Hence simple expressions for the magnetic coefficients are not easily extracted. In addition, insightful expansions are especially needed to support recent nanotechnology advancements, where all-dielectric devices exploit the strong magnetic and electric resonance of their elementary building blocks, such as spheres~\cite{Jahani2016,Garcia-Etxarri2011}.

In this work we propose an alternative system ansatz as a way to extract valuable physical information for the scattering resonant behavior of a small sphere. By studying expressions~(\ref{an}) and~(\ref{bn}) it becomes clear that a complicated zero/poles resonant scheme occurs for a given material-size combination. Notably, the Taylor series expansion of such resonating expressions might converge slowly, especially close to the poles. Moreover, coefficients~(\ref{an}) and~(\ref{bn}) are in a fractional function form, making clear that a system ansatz capable of describing the coefficients as a fractional set of rational functions could possibly reveal their zero/pole trends,  providing us with the necessary intuition about the scattering resonant behavior. Such system approximation covering all the above features is the Pad\'{e} approximant~\cite{baker1996pade}.

The Pad\'{e} approximant is a special type of rational fraction approximation~\cite{baker1996pade}, used particularly for the description of many physical problems where a complex resonant physical system is described and/or observed, such as in cosmology~\cite{Zunckel2008} or in quantum chromodynamics~\cite{Samuel1995}. This is due to its inherent ability to describe a function as a set of a rational polynomial functions, expanding each of the fractional terms in a power series polynomial~\cite{baker1996pade,Masjuan2014}.  

To our knowledge, similar expressions were first given by Wiscombe~\cite{Wiscombe1980} and recently in ~\cite{Alam2006,Ambrosio2012} giving a numerically efficient way for evaluating the Mie coefficients in the small size parameter limit. However, these studies do not focus on either the resonant conditions or the physical mechanisms for either electric or magnetic resonances of a homogeneous sphere. An equivalent approximative procedure has been followed in~\cite{Tribelsky2006}; Mie coefficients were decomposed in a fractional $\frac{r}{r+is}$ form and a Taylor series expansion calculated for each $r$ and $s$ term. In this way the pole conditions are obtained for the electric multipoles at the plasmonic regime, revealing peculiarities on the scattering behavior of the plasmonic sphere. 

However, the Pad\'e approximants of a fractional function, such as the Mie coefficients, are not equal with the fraction of two Taylor expanded functions, especially for low order approximations. This is mostly attributed to the inherent ability of the Pad\'e series to converge  quickly, especially close to singular points such as the poles of a system~\cite{baker1996pade}. In this way the expanded terms are simple and compact, allowing at the same time a much easier physical interpretation of the scattering mechanisms enabled. 


Let us begin with a magnetically transparent sphere ($\mu_c=1$). The lowest available approximants read,
\begin{equation}\label{a1P_dielectric}
 a_1^{[3/2]}=-i\frac{2}{3}\frac{\varepsilon_c-1}{\varepsilon_c+2}\frac{x^3}{(1-\frac{3}{5}\frac{\varepsilon_c-2}{\varepsilon_c+2}x^2)}
\end{equation}
\begin{equation}\label{b1P_dielectric}
  b_1^{[5/2]}=-i\frac{\varepsilon_c-1}{45}\frac{x^5}{1+\frac{1}{21}(5-2\varepsilon_c)x^2}
\end{equation}
where $b_1^P$ term is a [5/2] and $a_1^P$ a [3/2] Pad\'{e} expression, respectively. Key findings of this work can be extracted by carefully analyzing their pole behavior.

Starting with the electric coefficient (Eq.~(\ref{a1P_dielectric})), the Taylor expanded pole condition reads
\begin{equation}\label{TA}
 \varepsilon_{a_1}^{[3/2]}=-2-\frac{12}{5}x^2+O[x]^4
\end{equation}
where $O[x]^4$ denotes the truncated terms. Notice that the superscript denotes the used [L/M] Pad\'{e} approximant, while the subscript denotes the corresponding Mie term.

For vanishingly small size parameter values, Eq.~(\ref{TA}) yields to $\varepsilon_c\rightarrow-2$. This can be recognized as the electrostatic polarization enhancement condition~\cite{sihvola2006character} (or Fr\"{o}hlich frequency~\cite{bohren2008absorption}), obtained also by the Taylor series expansion in Eq.~(\ref{a1Taylor2}). However, by inspecting Eq.~(\ref{a1P_dielectric}) one notices that this is not a sufficient condition for the system to resonate: the resonant behavior is dependent on how this limit is approached. For instance when ($x\rightarrow0$) the $a_1$ coefficient goes to zero; the limiting case where a small sphere approaches $\varepsilon_c\rightarrow-2$ gives a finite value for the expression \ref{a1P_dielectric}, i.e., $i\frac{5}{6}x$. Therefore, condition $\varepsilon_c=-2$ is not a system pole, but rather an asymptotic limit derived from the electrostatic case~\cite{Mei2013}.

The above expressions verify already known results that can be found in textbooks, i.e.,~\cite{bohren2008absorption} (Ch.12, p.329), where a rough Taylor approximation of the quasistatic polarizability has been used. A similar but less accurate condition is extracted in~\cite{Tretyakov2014}, while a generalization for higher electric multipoles is given in~\cite{Tribelsky2011}. Notice that a comparison between condition in Eq.~(\ref{TA}) and the obtained values by Eq.~(\ref{an}) (for $n=1$) reveals that the relative error is less that $0.1\%$ for size parameters up to $x=0.4$.

\begin{figure}
\centering
 \includegraphics[width=0.5\textwidth]{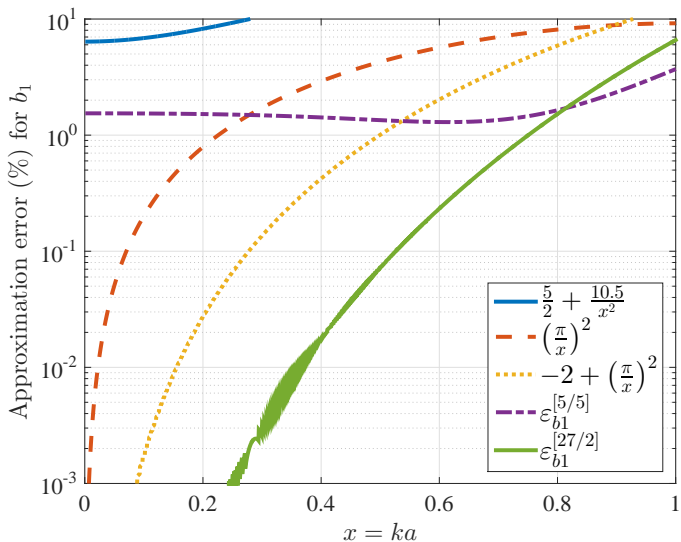}
         \caption{Approximation error ($\%$) between conditions of Eq.~(\ref{b1P_pole}) (blue solid), second term of Eq.~(\ref{b1_272}) (red dashed), the first two terms of Eq.~(\ref{b1_272}) (yellow dotted), Eq.~(\ref{b1_55}) (purple dot-dashed), and Eq.~(\ref{b1_272}) (green-solid) with respect to the values obtained from the Mie coefficients of Eq.~(\ref{bn}) for the first magnetic dipole. Equation.~(\ref{b1_272}) gives less than $1\%$ error for values up to $x=0.8$.}
        \label{comp}
\end{figure}

The next step is to expand our study for the case of the magnetic resonances. The Pad\'{e} expansion of the first magnetic Mie term (Eq.~(\ref{b1P_dielectric})) exhibits a pole with the following condition:
\begin{equation}\label{b1P_pole}
  \varepsilon_{b_1}^{[5/2]}=\frac{5}{2}+\frac{10.5}{x^2}+O[x]^2
\end{equation}
This resonance follows an inverse square size dependence, explaining that for very small size parameters the first magnetic resonance is hidden in the far positive permittivity axis. Many qualitative differences are derived by comparing the electric (Eq.~(\ref{TA}), $\varepsilon_c<0$,~\cite{Fan2014}) and magnetic (Eq.~(\ref{b1P_pole}), $\varepsilon_c>0$,~\cite{Geffrin2012}) pole conditions; plasmonic resonances (Eq.~(\ref{TA})) are less sensitive to size parameter and appear in material with smaller permittivity contrast. Although Eq.~(\ref{b1P_pole}) gives a poor approximation, having less $10\%$ error only for sizes up to $x<0.3$ (Fig~\ref{comp}), it can still predict the general resonant trend of the magnetic $b_1$ coefficient. 

Up to this point some simple rules regarding the first electric and magnetic dipole resonances have been derived. Arguably, expressions~(\ref{a1P_dielectric}) and~(\ref{b1P_dielectric}) are purely imaginary quantities for the lossless case, thus energy conservation is violated--see~\cite{capolino2009theory} (Ch.8, p.6). This can be immediately restored by introducing higher Pad\'{e} approximants, e.g. [3/3] for $a_1$ and [5/5] for $b_1$, viz.,
\begin{equation}\label{a1_55}
 a_1^{[3/3]}=-i\frac{2}{3}\frac{\varepsilon_c-1}{\varepsilon_c+2}\frac{x^3}{\left(1-\frac{3}{5}\frac{\varepsilon_c-2}{\varepsilon_c+2}x^2-i\frac{2}{3}\frac{\varepsilon_c-1}{\varepsilon_c+2}x^3\right)}
\end{equation}
\begin{equation}\label{b1_55_c}
\begin{aligned}
 &b_1^{[5/5]}=-i\frac{\varepsilon_c-1}{45}
 \\
 &\frac{x^5}{\left(1+\frac{1}{21}(5-2\varepsilon_c)x^2+[x]^4-i\frac{1}{45}(\varepsilon_c-1)x^5\right)}
 \end{aligned}
\end{equation}
where $[x]^4=-\frac{\varepsilon_c^2+100\varepsilon_c-125}{2205}x^4$.

Terms found in the denominator of Eq.~(\ref{a1_55}) and~(\ref{b1_55_c}) can be categorized into two types: real terms describing the dynamic depolarization effects~\cite{Meier1983}, and imaginary terms representing the \emph{radiative damping} effects~\cite{DeVries1998}, respectively. Notice that for $a_1$ and $b_1$ the [3/3] and [5/5] Pad\'{e} approximants are the lowest order approximants with an imaginary term in their denominator; these terms appear also in the Taylor expansion--see Eq.~(\ref{a1Taylor2}), and~(\ref{b1Taylor2}). In a sense, the form and order of the radiative damping term is known once the first Taylor term is calculated.

Following the previous analysis, Eq.~(\ref{a1_55}) gives the following pole condition  
\begin{equation}\label{eps_comp}
 \varepsilon_{a_1}^{[3/3]}=-2-\frac{12}{5}x^2-2ix^3+...
\end{equation}
where the imaginary term appeared reveals that a complex pole is expected, even for the lossless case due to radiative damping effects~\cite{Osipov2015}. This expression elucidates the fact that Mie coefficients exhibit resonances at complex frequencies, known also as \emph{natural frequencies}~\cite{stratton2007electromagnetic}, offering an equivalent definition for these frequencies and a clear interpretation from a material point of view.

Let us continue by assuming material losses, i.e.,$\varepsilon_c(\omega)=\varepsilon'(\omega)+i\varepsilon^{\prime\prime}(\omega)$. The estimated imaginary part of Eq.~(\ref{eps_comp}) reveals another interesting fact: the amount of dissipative losses required for maximum absorption efficiency are dictated by the amount of the radiative damping losses. This can be understood as a matching process between the two mechanisms involved, i.e., radiative damping and material dissipative losses~\cite{Tribelsky2011,Tretyakov2014}. This interchange raises a series of interesting and peculiar phenomena affecting the overall absorptive behavior of a sphere~\cite{LukYanchuk2012}. Therefore, the required material losses for maximum absorption for the first plasmonic resonance can be approximately estimated to be
\begin{equation}\label{a1_cont}
 \varepsilon^{\prime\prime}(\omega)\approx2x^3
\end{equation}

Similarly, the pole of Eq.~(\ref{b1_55}), rounded to the fourth decimal digit, reads

\begin{equation}\label{b1_55}
\begin{aligned}
 \varepsilon_{b_1}^{[5/5]}= &-2.0743+\left(\frac{3.1614}{x}\right)^2-1.4263x^2+...\\
		& -i2x\left(1.0673-0.7721x^2+0.4604x^4\right)+...
\end{aligned}
 \end{equation}
where up to $x^2$ dynamic depolarization (real part) and $x^5$ radiative damping terms (imaginary part) are included, respectively. A first rough approximation for maximum absorption reads (see~\cite{Osipov2015})
\begin{equation}\label{b1_cont}
 \varepsilon^{\prime\prime}(\omega)\approx2x
\end{equation}
easily derived by neglecting the higher order imaginary terms of Eq.~(\ref{b1_55}). 

\begin{figure}[!]
\centering
 \includegraphics[width=0.5\textwidth]{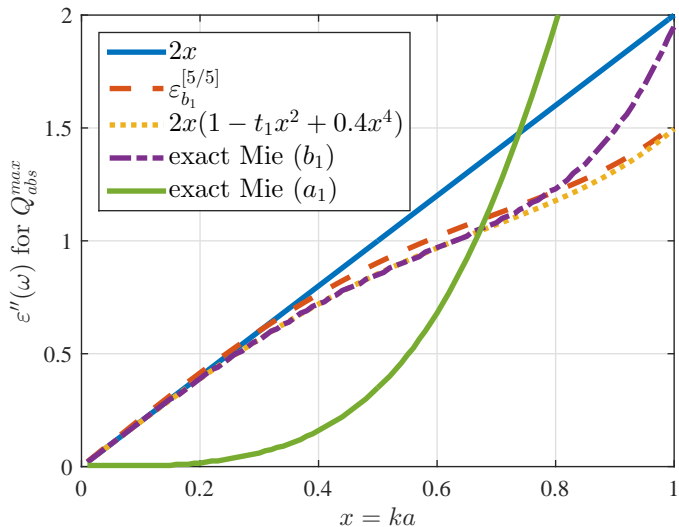}
         \caption{Calculated $\varepsilon^{\prime\prime}(\omega)$ values as a function of size parameter for maximum absorption for the first magnetic term ($b_1$). Blue solid line: the linear trend of Eq.~(\ref{b1_cont}). Red dashed line: the estimation given by Eq.~(\ref{b1_55}). Yellow dotted line: the estimation given by Eq.~(\ref{b1_272}); $t_1$ value can be found in Table~I. Purple dot-dashed line illustrates the absorption trend extracted from the exact Mie coefficient (Eq.~(\ref{bn})). The corresponding absorption maximum for the first electric Mie term (Eq.~(\ref{an}) green solid) exhibiting a monotonous, $2x^3$ distribution. The absorption plateau is visible for $0.4<x<0.8$. }
        \label{Abs}
\end{figure}

Consider now the estimated radiative damping terms of Eq.~(\ref{b1_55}). These, non-trivial terms predict a non-linear trend for the maximum absorption curve. Indeed, as can be seen in Fig.~\ref{Abs}, a maximum absorption plateau is observed around $x=0.6$. We characterize this predicted plateau as a manifestation of a \emph{self-regulating} radiative damping process. This is justified from the point that radiative damping is an intrinsic mechanism, affected only by the size characteristics in the small size limit, with immediate effects on the scattering and absorption processes. Hence, this non-linear trend extracted in Eq.~(\ref{b1_55}) reflects the ability of a sphere to exhibit different qualitative radiative damping behavior of these two types of resonances. Note that similar effects are not observed for the electric plasmonic resonances, where the absorption maximum curve is strictly monotonous for size parameters up to $x<1$, as can be observed in Fig.~\ref{Abs} (green curve).

So far the proposed approximation has offered simple and compact expansions for both Mie coefficients. These features are mostly derived using the lowest Pad\'{e} approximants. In order to increase the accuracy of the extracted conditions of the magnetic resonances, higher order approximants are needed. For instance a [27/2] expansion of $b_1$ yields to the following pole condition
\begin{equation}\label{b1_272}
\begin{aligned}
\varepsilon_{b_1}^{[27/2]}=&-2+\left(\frac{\pi}{x}\right)^2+1.696x^2-1.1232x^4+...\\
			   &-i2x(1-0.696x^2+0.4447x^4)+...
\end{aligned}
\end{equation}
A comparison between the estimated approximations (Fig.~\ref{comp}) reveals that Eq.~(\ref{b1_272}) gives highly accurate results, while the accuracy is still high by including only the first two terms of Eq.~(\ref{b1_272}).
Note that in coefficient expansions like (\ref{b1_55}) or higher, more than one poles are predicted. For a given physical system some of the predicted poles may not be observed directly, making their physical interpretation a difficult task~\cite{Masjuan2014}. Additionally, some of the poles may coincide with system zeros, hence their effects are canceled. Our study is restricted only for physically observable poles, verified through the analytical Mie solution.

We conclude our discussion by generalizing the extracted results for the case of higher order magnetic resonances. A simple pattern regarding their resonant condition can be extracted by carefully analyzing the higher order Pad\'{e} approximants and their poles, viz.,

\begin{equation}\label{general}
 \varepsilon_{bn}=-\frac{2}{2n-1}+\left(\frac{c_n}{x}\right)^2-i\frac{2}{\left[(2n-1)!!\right]^2}x^{2n-1}\left(1-t_nx^2\right)
\end{equation}
where $n=1,2,3,...$ is the order of each mode; $c_n$ and $t_n$ values are given in the following table.
\begin{table}[h]
\centering
\caption{Coefficients of Eq.~(\ref{general})}
\label{t1}
\begin{tabular}{l*{5}{c}r}
$n$              & 1 & 2 & 3 & 4 & 5 \\
\hline
\\
$c_n$ 		& $\pi$ & 4.4934 & 5.7634 & 6.9879 & 8.1825 \\
$t_n$            & 0.6960 & 0.2507 & 0.1578 & 0.1165 & 0.0928
\end{tabular}
\label{table}
\end{table}

The real part of Eq.~(\ref{general}) is a rule for the resonant position; the absorption maximum for each mode can be approximately described by the imaginary part. Notice that $c_n$ coefficients follow the order of the first zero of spherical Bessel function ($j_{n-1}\left(c_{n}\right)=0$).

{A new, Pad\'{e} approximant-based system ansatz has been introduced for the Mie coefficients, describing the scattering and absorptive mechanisms in a homogeneous sphere. Novel aspects and accurate trends for the magnetic multipoles resonant locations were revealed, while simple and compact coefficient expansions were introduced. This perspective can be further generalized for dispersive material models~\cite{kreibig1995optical}, inhomogeneities~\cite{Monticone2014}, anisotropies~\cite{Wallen2015}, or other geometries~\cite{Ruan2010}, revealing potentially interesting and unknown radiation/light scattering phenomena. Consequently, new design guidelines will emerge regarding the scattering and absorptive functionalities of single, canonical shaped scatterers.}

This work is supported by the Aalto Energy Efficiency Program (EXPECTS project) and the Aalto ELEC Doctoral School scholarship.

\bibliography{prb}
%
%
%
%
%
%
%
%
%
%

\end{document}